\documentclass[a4paper,11pt]{article}
\usepackage{pos}

\newcommand{\beqn}{\begin{eqnarray}}
\newcommand{\eeqn}{\end{eqnarray}}
\newcommand{\beqs}{\begin{subequations}}
\newcommand{\eeqs}{\end{subequations}\\[-2mm]\noindent}
\newcommand{\eq}[1]{(\ref{#1})}

\newcommand{\bs}{\boldsymbol}

\allowdisplaybreaks

\definecolor{brickred}{rgb}{0.8, 0.25, 0.33}
\definecolor{macouleur}{RGB}{105,150,150}

\usepackage{color}
\definecolor{purple}{rgb}{0.8,0,0.6}

\title{The Casimir Effect in (3+1)-dimensional lattice Yang-Mills theory at finite temperature: the unexpected universality of quarkiton and glueton boundary states}
\ShortTitle{The Casimir Effect in Yang-Mills theory}

\author[a]{Maxim N. Chernodub}
\author[b]{Vladimir A. Goy}
\author[b,c]{Alexander V. Molochkov}
\author[b]{Konstantin R. Pak}
\author*[b]{Alexey S. Tanashkin}

\affiliation[a]{Institut Denis Poisson UMR 7013, Universit\'e de Tours, \\
20 Avenue Monge - Parc Grandmont, 37200 Tours, France}

\affiliation[b]{Pacific Quantum Center, Far Eastern Federal University,\\
10 Ajax Bay, Russky Island, 690922 Vladivostok, Russia}

\affiliation[c]{Beijing Institute of Mathematical Sciences and Applications,\\
No. 544, Hefangkou Village Huaibei Town, Huairou District Beijing 101408, China }

\emailAdd{tanashkin.alexey@gmail.com}

\abstract{In our earlier work on the Casimir effect in $(3+1)$-dimensional Yang-Mills theory, we identified two novel nonperturbative states arising in QCD with boundaries: the glueton and the quarkiton. The glueton, or ``gluon exciton'', is a colorless bound state formed by gluons interacting with their negatively colored images in a chromometallic mirror. The quarkiton, or ``quark exciton'', is a meson-like state comprising a heavy quark attracted to its image through the mirror. In this study, we extend our analysis to finite temperatures near the deconfinement phase transition $(T \approx 0.78 T_c)$, where we observe a linear potential between a color-neutral chromometallic mirror and a heavy test quark. Our result suggests that the quarkiton state can have a physical relevance since mirrors for photons and, presumably, gluons can be realized in field theories as domain-wall solutions. Furthermore, we find a striking universality: the ratio of the glueton mass to the bulk $0^{++}$ glueball mass—defining the bulk mass gap—matches the ratio of the quarkiton string tension to the string tension between quark and anti-quark in the absence of the mirror, with a value ${\cal R} = 0.294(11)$.}

\FullConference{The XVIth Quark Confinement and the Hadron Spectrum Conference (QCHSC24)\\
 19-24 August, 2024\\
 Cairns Convention Centre, Cairns, Queensland, Australia\\}

\begin{document}
\maketitle

\section{Introduction}
The Casimir effect manifests itself as the emergence of a force between charge-neutral physical macroscopic objects~\cite{Casimir1948, Mostepanenko1988}. The observation of this force in several experiments~\cite{Lamoreaux1997, Mohideen1998} is generally regarded as the confirmation of the existence of the zero-point energy and the associated quantum vacuum fluctuations~\cite{Milonni1994, Milton2001, Bordag2009}. In field theories, the reflective mirrors may be realized as domain-wall solutions~\cite{Battye:2021dyq}, which opens up a possibility to study the Casimir effect in particle physics.

In a series of papers~\cite{Chernodub2016, Chernodub2017, Chernodub2017a, Chernodub2018, Chernodub2022, Chernodub2023}, we focused on non-perturbative aspects of the Casimir effect. We demonstrated that the Casimir effect not only generates the forces between (metallic) boundaries but also leads to a restructuring of the vacuum in the space between them, implying the effect of the boundaries is not restricted only to a simple modification of the spectrum of vacuum fluctuations~\cite{Chernodub2017, Chernodub2017a, Chernodub2018, Chernodub2022, Chernodub2023}. The summary of these results can be found in reviews~\cite{Chernodub2019, Molochkov2023}. 

In our recent paper~\cite{Chernodub2023}, we proposed the existence of two new excitations in QCD, a glueton and a quarkiton, that appear as novel boundary states in a non-Abelian gauge theory. The glueton is the nonperturbative state of a gluon bounded to its negative image in the chromometallic mirror. We have shown that in pure SU(3) Yang-Mills theory, the mass of the glueton is more than 3 times lighter than the mass of the ground state of $0^{++}$ glueball~\cite{Chernodub2023}. For the quarkiton, which is a bound state of a quark with its mirror image, we have not obtained the precise parameters of the potential, and, therefore, we restricted ourselves only to qualitative considerations, leaving a quantitative analysis for future work. The glueton and the quarkiton boundary states may presumably be related to the dynamical edge modes in vector gauge theories discussed recently in Refs.~\cite{Ball:2024gti, Ball:2024hqe, Canfora:2024awy}.

In the present paper, we report preliminary results on the properties of the quarkiton by analyzing how the free energy of heavy quarks depends on the distance to the mirror at finite temperatures. At zero temperature, this method proves to be ineffective due to the ill-defined nature of the Polyakov loop in the $T \to 0$ limit. To circumvent this technical problem, we set the finite temperature at $T \approx 0.78 T_c$, which allows us to determine the parameters of the potential between a heavy quark and the mirror. 

\section{Yang-Mills theory and Casimir boundary conditions on the lattice}
The action of continuous Yang-Mills theory in $(3+1)d$ Minkowski spacetime has the form
\begin{equation}
    S = - \frac{1}{4} \int d^4 x \, F_{\mu\nu}^a F^{a,\mu\nu} \,.
    \label{eq_S_continuum}
\end{equation}
The chromometallic mirror, placed in the ($x_1,x_2$)-plane, can be modeled by the gauge-invariant non-Abelian Casimir boundary conditions:
\begin{equation}
	E^a_{\|}(x) {\biggl|}_{x\in {\cal S}} \! = B^a_{\perp}(x) {\biggl|}_{x\in {\cal S}} \! = 0, 
    \qquad\
    a = 1,\dots, N^2 - 1.
    \label{eq_Casimir_continuum}
\end{equation} 
These conditions imply that the tangential chromoelectric fields $E^a_i \equiv F^a_{0i}$ and normal chromomagnetic fields $B^a_i = (1/2) \varepsilon_{ijk} F^{a,jk}$ vanish at the surface ${\cal S}$. The boundary conditions~\eq{eq_Casimir_continuum} are identical, up to the color index $a = 1,\dots, N^2-1$, to the conditions imposed on the Abelian electromagnetic field at the surface of a perfectly conducting metal in electrodynamics. The latter corresponds to an ideal mirror for photon fields. The conditions in  Eq.~\eq{eq_Casimir_continuum} extend this construction to non-Abelian fields, thus setting up a chromometallic mirror plate for gluons.

The particularities of the formulation of the Casimir boundary conditions on the lattice have been thoroughly discussed in our previous papers~\cite{Chernodub2022, Chernodub2023}. Below, we will briefly recall certain essential points of the construction, referring the interested reader to Ref.~\cite{Chernodub2016} for more details.

The Wilson formulation of the lattice Yang-Mills action~\eq{eq_S_continuum} is given by a sum over lattice plaquettes $P {\equiv} P_{n,\mu\nu} {=} \{n,\mu\nu\}$:
\beqn
	 S = \beta\sum_{P} \left(1 - {\mathcal P}_P \right), \qquad
      {\mathcal P}_P = \frac{1}{3}\mathrm{ReTr}\,U_P,\quad
\label{eq_action}
\eeqn
where $\mu$ and $\nu$ label the directions of the axis, $n$ denotes a site of a $4d$ Euclidean lattice, and $\beta = 6/g^2$ is the lattice coupling of SU(3) gauge theory. In the continuum limit, the lattice spacing vanishes $a\to 0$, the lattice plaquette $U_{\mu\nu}(n) = U_\mu(n)U_\nu(n+\hat\mu)U^\dag_\mu(n+\hat\nu) U^\dag_\nu(n) = \exp(i a^2 F_{\mu\nu}(n) + \mathcal{O}(a^3))$ reduces to the continuum field-strength tensor~$F_{\mu\nu}$, and the lattice action~\eq{eq_action} becomes a Euclidean version of continuum Yang-Mills action~\eq{eq_S_continuum}. 

The Casimir boundary conditions~\eq{eq_Casimir_continuum} in the Euclidean lattice formulation are achieved by promoting the lattice coupling in Eq.~\eq{eq_action} to a plaquette-dependent quantity $\beta \to \beta_P$. Here, one sets $\beta_P = \lambda \beta$ if the plaquette $P$ either touches or belongs to the world hypersurface spanned by the surface ${\cal S}$ and $\beta_P = \beta$ otherwise~\cite{Chernodub2016}. The quantity $\lambda$ plays the role of a Lagrange multiplier, which, in the limit $\lambda \to \infty$, enforces the lattice version of the Casimir boundary conditions~\eq{eq_Casimir_continuum}. 

The physical temperature is inversely proportional to the temporal extension of the lattice, $N_t$, where $N_s \gg N_t$ is the size of the lattice in the spatial directions~\cite{Gattringer2010}:
\beqn
    T = \frac{1}{N_t a(\beta)}\,.
    \label{eq_temp_phys}
\eeqn

We study the potential between the heavy quark and the chromometallic mirror at various lattice spacings $a = a(\beta)$ to make sure that our results are not significantly influenced by lattice artifacts associated with lattice discretization and finite volume effects. To maintain consistency of our approach, we keep the physical temperature $T$, Eq.~\eqref{eq_temp_phys}, constant while the other parameters of the lattice, namely the temporal dimension of the lattice $N_t$ and the lattice constant $\beta$, vary.

We performed our simulations at the fixed temperature $T = 0.5\sqrt\sigma \approx 0.78T_c$, which allows us to stay deeply in the confinement phase --- a necessary requirement to get insight on the $T=0$ QCD bound states--- while also keeping the value of lattice coupling $\beta$ in a physical range. The corresponding set of lattice and physical parameters is given in Table~\ref{table_param}.

\begin{table}[h!]
\centering
\begin{tabular}{| c | c | c | c | c | c |}
\hline
$N_t$ & $\beta$ & $a\sqrt\sigma$ & $\beta_c$ & $a\sqrt\sigma_c$ & $T/T_c$\\
\hline
5 & 5.6924 & 0.3999 & 5.8000 & 0.3176 & 0.7918 \\
\hline
6 & 5.7762 & 0.3333 & 5.8941 & 0.2612 & 0.7837 \\
\hline
7 & 5.8492 & 0.2856 & 5.9800 & 0.2236 & 0.7829 \\
\hline
8 & 5.9174 & 0.2499 & 6.0625 & 0.1947 & 0.7791 \\
\hline
\end{tabular}
\caption{The values of lattice sizes and the coupling constants used in our calculations. The data for the lattice string tension $a\sqrt\sigma$ is taken from Ref.~\cite{Athenodorou2020} while the critical values are taken from Ref.~\cite{Cardoso2012}. Certain values in this table are obtained using an interpolation procedure by cubic splines.}
\label{table_param}
\end{table}

\section{Free energy of heavy quark}
In our first work~\cite{Chernodub2023}, we studied the free energy  $F_{Q|}$ of heavy quark ``$Q$'' as a function of its distance $d$ to the mirror ``$|$''. While that calculation has been formally performed at zero temperature, the physical extension of the lattice in the temporal direction $L_t$ has been taken finite, $L_t \equiv N_t a \leqslant L_s \equiv N_s a$. The free energy of a heavy quark is related to the expectation value of the Polyakov loop near the mirror boundary:
\beqn
    \langle P_{\bs{x}} \rangle_{|}(d) = \exp\{- L_t F^{\rm lat}_{Q|}(d)\}\,.
    \label{eq_free_en}
\eeqn
This quantity, strictly speaking, is not a suitable zero-temperature observable in the thermodynamic limit because the latter implies $L_t \to \infty$ thus rendering the Polyakov loop~\eqref{eq_free_en} exponentially vanishing. Therefore, in our previous work, we had to choose a rather small lattice $12^4$ to get a qualitative picture of the effect of the boundary on the free energy of a heavy quark. Given the small volume of the lattice, the quantitative information that we obtained in our previous study was rather limited.

Despite the limitations posed by the finiteness of the lattice volume, we found the exciting signatures that the free energy of a heavy quark $F^{\rm lat}_{Q|}$ linearly rises as the function of the distance to the mirror~\cite{Chernodub2023}. This observation shows the existence of the attraction between the quark and the neutral mirror, which is interpreted as a result of the emergence of a confining string between these objects. In this work, we repeat the same calculation at a finite temperature and confirm our previous result with a much higher accuracy. 

\begin{figure}[ht]
\centering
  \includegraphics[width=0.7\linewidth]{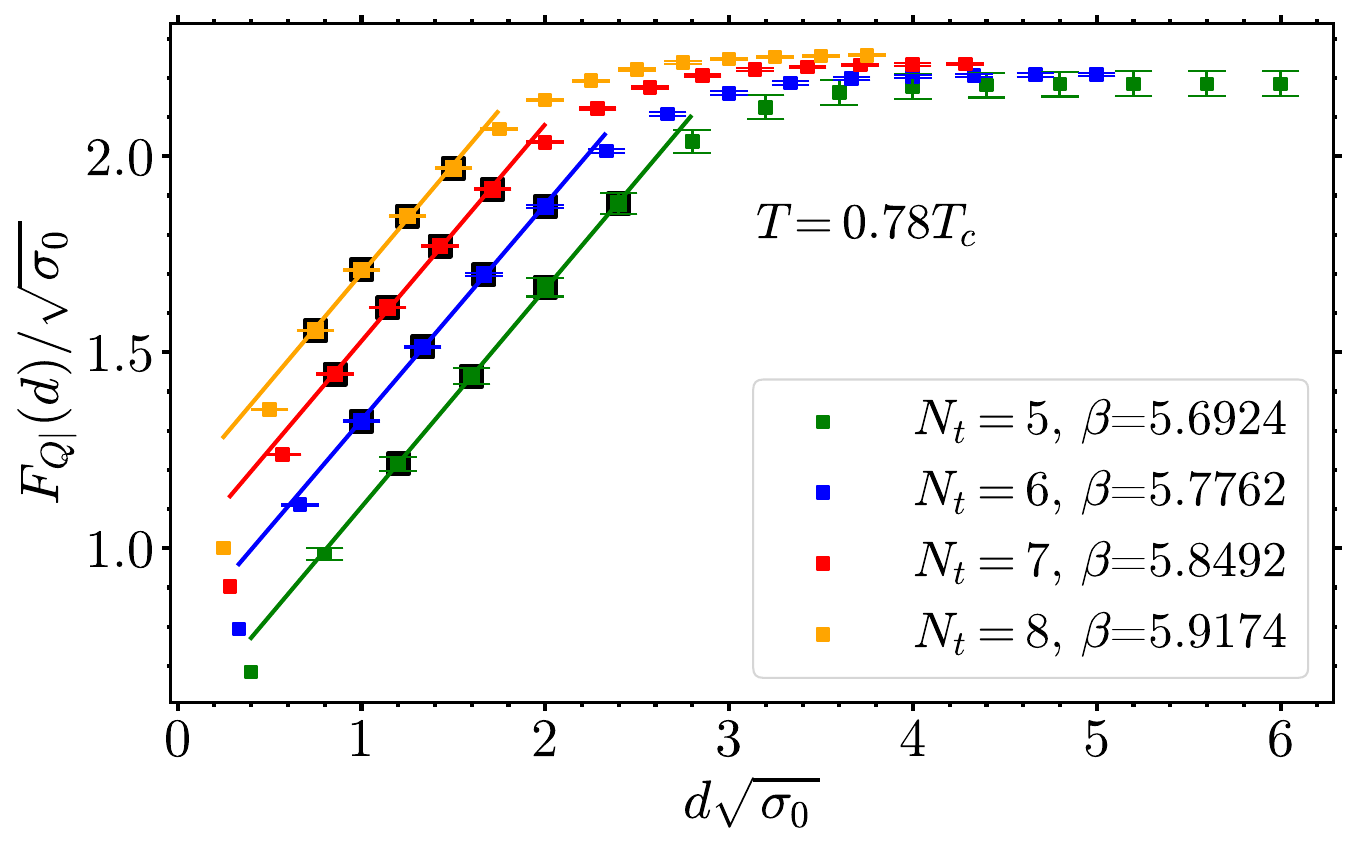}
  \caption{Unrenormalized free energy of heavy quark as a function of distance to the mirror in physical units at a fixed temperature.}
  \label{fig_FE_fixT_phys}
\end{figure}

Figure~\ref{fig_FE_fixT_phys} shows the behavior of the non-renormalized quark free energy $F^{\rm lat}_{Q|}$ as a function of the distance of the static heavy quark $Q$ to the plane $d$. All quantities are expressed in units of the physical string tension at zero temperature $\sqrt\sigma_0$.
A linear slope in the free energy $F^{\rm lat}_{Q|}$ for each value $\beta$ is clearly recognizable. To guide the eye, these linear segments of the potentials are shown explicitly in Fig.~\ref{fig_FE_fixT_phys} by linear functions.

The heavy quark free energy possesses two types of artifacts related to infrared and ultraviolet cut-offs imposed in our calculations. At large distances from the mirror boundary, the free energy flattens due to the finite volume of the system, so that the linear slope does not extend to spatial infinity. In the following, we exclude the artificial large-distance region from our discussion.

\begin{figure}[ht]
\centering
  \includegraphics[width=0.7\linewidth]{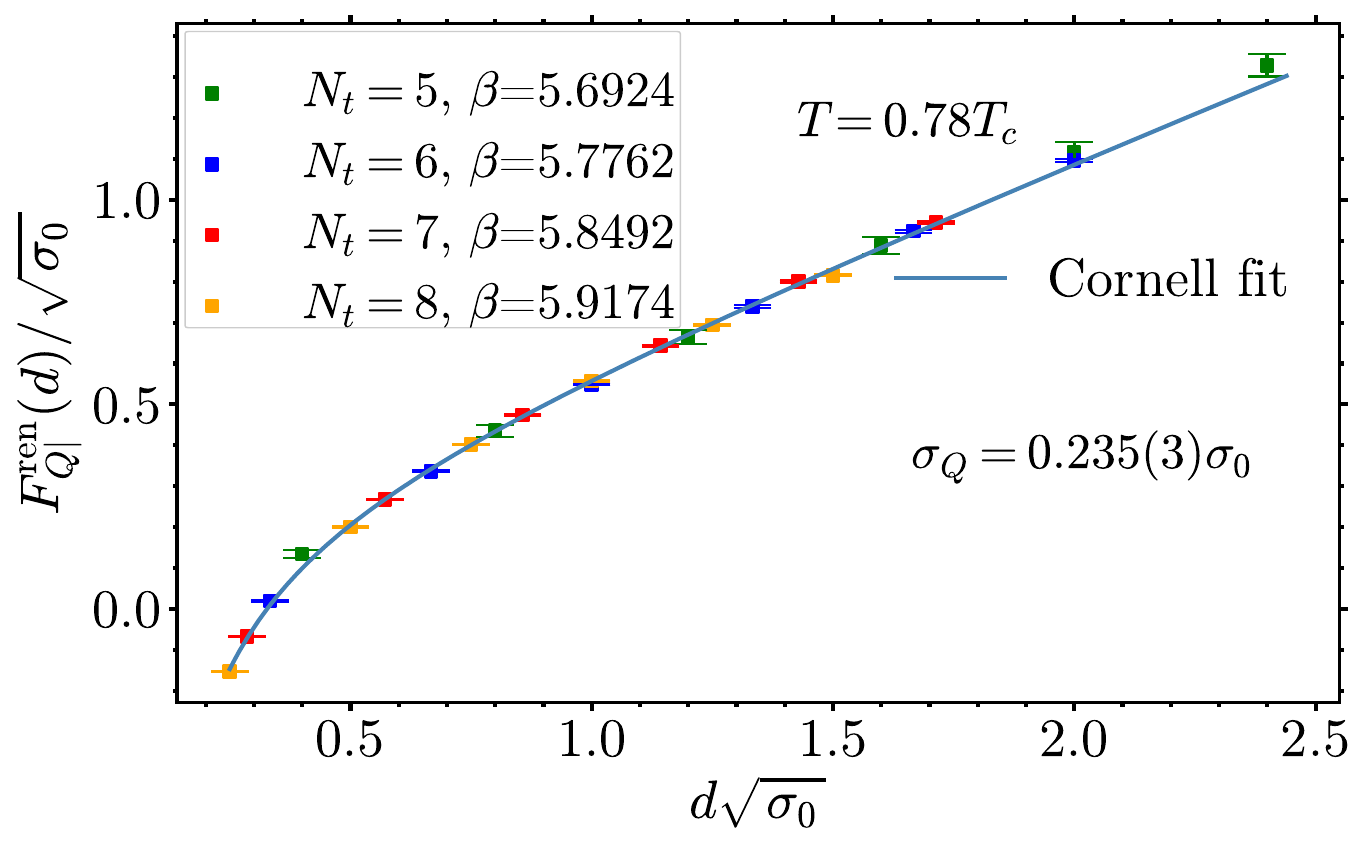}
  \caption{The renormalized free energy of heavy quark as a function of distance to the mirror in the physical units at the fixed temperature $T = 0.78 T_c$. The fit by the Cornell potential~\eqref{eq_cornell_fit} is shown by the solid curve.}
  \label{fig_scaling_final}
\end{figure}

The ultraviolet renormalization additively renormalizes the free energy, shifting the lattice data by a finite quantity, $F_{Q|}^\mathrm{lat}(d,\beta) = F_{Q|}^\mathrm{ren}(d) + \Delta F(\beta)$, where $F_{Q|}^\mathrm{ren}(d)$ is the physical (``renormalized'') free energy of a heavy quark that does not depend on the lattice coupling $\beta$. We found that, within our numerical accuracy, the lattice free energy of the heavy quark can be renormalized by additively shifting the lattice free energy by the simple linear function $\Delta F(\beta) = a_0 + a_1 \beta$, with parameters $a_0 = -14.72(17)$ and $a_1 = 2.68(3)$. After subtracting this contribution from the lattice data and excluding the flattening long-distance tails, we finally get the renormalized free energy $F_{Q|}^\mathrm{ren}(d)$, as shown in Fig.~\ref{fig_scaling_final}.

The data for the renormalized free energy points perfectly collapse to the single curve, thus indicating the independence of our data both on the ultraviolet lattice cutoff $a = a (\beta)$ and the spatial lattice size $L_s = N_s a$. Moreover, the renormalized free energy can be nicely fitted by the Cornell potential,
\beqn
    F_{Q|}^\mathrm{ren}(d) = - \frac{\alpha_{Q}}{d} + 2 \sigma_{Q} d + F_0\,.
    \label{eq_cornell_fit}
\eeqn
The first term in this expression represents the perturbative Coulomb ($\propto 1/d$) potential which appears due to a gluon exchange between the heavy quark and its image in the mirror. This term dominates at small distances. Our fit gives $\alpha_{Q} = 0.117(2)$ for the dimensionless prefactor. 

The second term in Eq.~\eqref{eq_cornell_fit} represents the non-perturbative linear ($\propto d$) growth of the free energy due to the formation of the confining string between the heavy quark and the neutral mirror. This term is important at large distances. The coefficient in the second term comes with the additional prefactor ``2'', which represents the fact that the quarkiton is a state of quark and its negative image in a chromometallic mirror. The image appears at the distance $d$ behind the mirror antiquark $\bar Q$, so that the actual distance from the quark to its mirror image, $d_{Q \bar Q}$, is twice as big as the distance of the original heavy quark $Q$ to the mirror itself, $d_{Q \bar Q} = 2 d_{Q |} \equiv 2 d$. The best-fit value for the string tension in Eq.~\eqref{eq_cornell_fit} is:
\beqn
    \sigma_{Q} = 0.235(3)\sigma_0
    \label{eq_sigma_qton}
\eeqn
The last term in Eq.~\eqref{eq_cornell_fit} gives us an inessential constant contribution with $F_0 = 0.204(9) \sqrt{\sigma_0}$.

The appearance of the linear potential between an isolated quark and a neutral mirror, shown in Fig.~\ref{fig_scaling_final} and fitted by the Cornell potential~\eqref{eq_cornell_fit}, provides us with the compelling evidence of the linear attraction of probe heavy-quark color charge to the mirror. This result shows that the quark is indeed confined to the chromoelectric mirror, thus supporting our previous conclusion on the formation of the non-perturbative quarkiton boundary state.

\section{The string tension of quarkiton}

The coefficient $\sigma_Q$ at the linear term in equation~\eq{eq_cornell_fit} can be naturally interpreted as the string tension of the quarkiton at temperature $T = 0.78 T_c$. It is interesting to compare the obtained result with the string tension $\sigma_{Q \bar Q}$ of the quark-antiquark pair at the same temperature. The value of $\sigma_{Q \bar Q}$ was obtained from the correlator of two Polyakov loops separated by distance $l$ in the absence of the mirror:
\beqn
    \langle P_{\bs{x}}P_{\bs{x+r}} \rangle = \exp\{-L_TF_{Q \bar{Q}}(r)\}
    \label{eq_corr_qq}
\eeqn
The free energy $F_{Q\bar{Q}}(r)$ was fitted by the Cornell potential~\eq{eq_cornell_fit} with $d \equiv r$. The best fit result for the string tension,
\beqn
    \sigma = 0.80(1)\sigma_0
    \label{eq_sigma_T}
\eeqn
is in good agreement with the result of Ref.~\cite{Cardoso2012}, where the string tension $\sigma_{Q\bar Q} = 0.7943(57)\sigma_0$ has been obtained at $T=0.788T_c$. Therefore, the ratio of the quarkiton string tension~\eq{eq_sigma_qton} to the quark--anti-quark string tension~\eq{eq_sigma_T} is equal to
\beqn
    {\mathcal R}_{Q|} = \frac{\sigma_Q}{\sigma} = 0.294(11) \ \qquad {\rm [for\ the\ quarkiton]}\,.
    \label{eq_ratio_qton}
\eeqn
Thus, the string tension of the quarkiton boundary state is more than 3 times smaller than the tension of the string of the $Q\bar{Q}$ pair. 

It is interesting to compare these results with the bound states of gluons with the chromometallic mirror. In our previous work~\cite{Chernodub2023}, we calculated the mass of the gluonic boundary state, the glueton. This mass also turned out to be around 3 times smaller than the mass of $0^{++}$ glueball, with the latter quantity setting the mass gap in the bulk of the system. Therefore, both for quarks and gluons, the boundary states have fewer masses than the corresponding colorless bound states in the absence of the mirror. But what is really remarkable is the fact that the corresponding ratios for the glueton and the quarkiton are equal to each other with remarkably good precision. The calculated mass of glueton was $m_\mathrm{gt}=1.0(1)\sqrt\sigma_0$ and $m_{0^{++}}$ is $3.41(2)\sqrt\sigma_0$. The corresponding ratio for the glueton is
\beqn
    {\mathcal R}_{{\rm g}|} = \frac{m_\mathrm{gt}}{m_{0^{++}}}=0.293(29) \ \qquad {\rm [for\ the\ glueton]}\,.
    \label{eq_ratio_gton}
\eeqn
The numerical coincidence of these ratios, given in Eqs.~\eqref{eq_ratio_qton} and \eqref{eq_ratio_gton}, points out the unexpected property that there exists some general universal scaling for the QCD boundary states, regardless of their quark or gluon nature.

\section{Conclusion}
In this paper, we studied the free energy of heavy quarks as a function of the distance to the color-neutral chromometallic mirror. To facilitate the numerical simulations, we performed our calculations at a finite temperature $T\approx0.788T_c$ in the confining phase of Yang-Mills theory. We demonstrated that a heavy quark is linearly attracted to the neutral mirror, thus supporting our initial suggestion of the existence of quark boundary states, quarkitons, in QCD. We found that the free energy of a heavy quark near the mirror is perfectly described by the Cornell-type potential~\eqref{eq_cornell_fit} as a distance of the quark to the mirror. We also calculated the corresponding string tension~\eqref{eq_ratio_qton} and pointed out that the ratio ${\mathcal R}_{Q|}$ of the string tension in the quarkiton state near the mirror to the string tension of quark--anti-quark state in bulk~\eqref{eq_ratio_qton}, and the ratio ${\mathcal R}_{{\rm g}|}$ of the mass of the glueton state near the mirror to the mass of the glueball in bulk~\eqref{eq_ratio_gton}, are equal to each other with good numerical accuracy ${\mathcal R}_{Q|} \simeq {\mathcal R}_{{\rm g}|}$, thus demonstrating an unexpected universality that emerges in the QCD boundary states. 

\acknowledgments
The numerical simulations were performed at the computing cluster Vostok-1 of Far Eastern Federal University. The work of V.A.G., A.V.M., K.R.P., and A.S.T. was supported by Grant No. FZNS-2024-0002 of the Ministry of Science and Higher Education of Russia.

\bibliographystyle{apsrev4-1}
\bibliography{quarkiton.bib}

\end{document}